\documentclass[prd,twocolumn,showpacs,nofootinbib]{revtex4}
\usepackage{amsmath}

\def\half{\tfrac 12}
\def\asflat{asymptotically flat}
\def\sph{spherically symmetric}
\def\const{{\rm const}}
\def\d{\partial}
\def\eps{\varepsilon}

\begin{document}

\title{Quasi-black holes:
	general features and purely field configurations}
\author{K. A. Bronnikov}
\email{kb20@yandex.ru}
\affiliation{VNIIMS, 46 Ozyornaya St., Moscow 119361, Russia,}
\affiliation{Institute of Gravitation and Cosmology, PFUR,
	6 Miklukho-Maklaya Street, Moscow 117198, Russia,}
\affiliation{National Research Nuclear University ``MEPhI'' 
	(Moscow Engineering Physics Institute)}
\author{O. B. Zaslavskii}
\email{zaslav@ukr.net}
\affiliation{%Department of Physics and Technology, 
	Kharkov V.N. Karazin
	National University, 4 Svoboda Square, Kharkov 61022, Ukraine}
\affiliation{Institute of Mathematics and Mechanics, Kazan Federal
	University, 18 Kremlyovskaya St., Kazan 420008, Russia}

\begin{abstract}
  Objects that are on the threshold of forming the horizon but never
  collapse are called quasi-black holes (QBHs). We discuss the properties 
  of the general \sph\ QBH metric without addressing its material source, 
  including its limiting cases as the corresponding small parameter tends to   
  zero. We then show that QBHs can exist among self-gravitating configurations 
  of electromagnetic and dilatonic scalar fields without matter.
  These general results are illustrated by explicit examples of exact 
  solutions.
\end{abstract}

\maketitle

\section{Introduction}

  During the recent two decades, the zoo of relativistic objects has
  increased and now, in addition to stars, black holes (BHs), wormholes,
  boson stars etc., it also includes the so-called quasi-black holes (QBHs).
  These are objects which are close to forming an event horizon but do not
  form it. Their more rigorous definition and general properties are
  described in \cite{qbh}. There are different types of sources that provide
  physical realizations of QBHs. They naturally appear in the so-called
  Majumdar-Papapetrou (MP) systems \cite{m, p} made of extremely charged
  dust, where gravitational attraction is balanced by electric repulsion,
  $\rho_{e} = \pm \rho_{m}$ $\rho_{m}$ and $\rho_{e}$ are the mass and
  charge densities, respectively (see \cite{ext} and references therein).
  Generalized MP systems including a dilaton scalar field (dilatonic MP,
  or DMP systems) have been suggested as QBH sources in \cite{1, 2}.

  The goal of the present paper is twofold. First, we summarize and
  generalize the previous observations and describe the general properties of
  QBH metrics in the framework of general relativity, independently of their
  specific material source. Second, we show that QBHs can exist without
  (macroscopic) matter at all, as a result of combined action of
  gravitational, electromagnetic and scalar fields. It is worth noting that
  purely field QBHs have already been considered in \cite{lw1, lw2}, being
  built with the aid of non-Abelian gauge and Higgs fields (self-gravitating
  monopoles). The main results of these papers were based on numerical
  calculations, so it is not easy to control them. A similar behavior was
  found analytically in \cite{jmp} using thin shells made of extremal dust.

  Unlike that, we will deal with Abelian fields only, which enables us to
  trace the properties of QBHs analytically and in more detail. To the best
  of our knowledge, this is a first example where a QBH is (i) purely field
  (vacuum), (ii) built using Abelian fields.

  In Section II we will discuss the properties of the general QBH metric 
  without addressing its material source. Section III is devoted to purely 
  field QBH models, and Section IV is a conclusion.
  Throughout the paper, we use the units in which $G=c=1$.

%%%%%%%%%%%%%%%%%%%%%%%%%%%%%%%%%%%%%%%%%%%%
\section{General features of QBH metrics}

  In both MP and DMP systems, in the case of spherical symmetry
  the metric has the form \cite{m,p,1,2}
\begin{equation}
	ds^2  = A(x) dt^2
		- A(x)^{-1} (dx^2 + x^2 d\Omega^2),  		\label{met}
\end{equation}
  where $d\Omega^2 = d\theta^2 + \sin^2 \theta\,d\varphi^2 $ is the linear
  element on a unit sphere.

  It turns out that this particular form of the metric (\ref{met}) entails
  some important consequences which are insensitive to the details of a
  material source that supports the metric. To discuss them, in this section
  we will ``forget'' physical motivations for the metric (\ref{met}) and
  look at it from a purely geometric viewpoint.

  To begin with, the metric (\ref{met}) describes a BH if $A(x)$ has a
  regular zero at some value of $x$, $x = x_{\rm hor}$, corresponding to the
  event horizon. A simple inspection shows that this can only be achieved
  at finite values of the spherical radius $r(x) = x/\sqrt{A(x)}$ if and
  only if
\begin{equation}
	x_{\rm hor} =0, \ \ \ A(x) = x^2 A_1(x),    	    \label{hor}
\end{equation}
  with a smooth function $A_1(x)$ such that $A_1(0)\ne 0$.

  Further, the system admits the existence of QBHs: to be close to a BH,
  such a configuration should contain a region with small $A(x)$ but have
  a regular center. Introducing a small parameter $c$, we can describe the
  situation by presenting $A(x) \mapsto A(x,c)$ as a Taylor series
\begin{equation}
     	A(x,c) = A_0 (c) + \half A_2 (c) x^2 + \ldots,        \label{A_c}
\end{equation}
  where $A_0 (c) \to 0$ as $c \to 0$ while $A_2(0)$ is finite.
  Redefining the parameter $c$ so that $A_0 (c) = \half A_2(0) c^2$,
  we can put without loss of generality
\begin{equation}
	A(x, c) = \frac{x^2 + c^2}{f^2(x,c)}\ \ \ \Rightarrow\ \ \
	r(x, c) = \frac{x f(x,c)}{\sqrt{x^2 + c^2}},     	\label{A,r}
\end{equation}
  where $f(x,c)$ is a smooth function with a well-defined nonzero limit as
  $c\to 0$. 

  This reasoning is the same as was applied to charged dust configurations
  in \cite{qbh, ext}. Let us stress that in the general case,
  with the metric function (\ref{A,r}), the region where the
  ``redshift function'' $A(x,c)$ is small, is itself not small at all:
  assuming that $f(x,c) = O(1)$ and $c \ll 1$, it is easily shown that the
  radius $r(c)$ of the sphere $x=c$ is $f(c,c)/\sqrt{2} = O(1)$; the
  distance from the center to this sphere and the volume of the ball
  $x \leq c$ are also $O(1)$.

\subsection{Scales and regions in QBH spacetimes}

  We will assume that all metrics are asymptotically flat, so that
  $A \to 1$ as $x\to \infty$, although the QBH concept is meanginful even
  without this assumption.

  According to \cite{qbh} and \cite{mn}, one of the typical features of QBHs
  consists in that there are at least two regions with their own
  characteristic scales on which physical and geometric quantities like
  density, metric coefficients, etc.  may vary. If there is a sharp boundary
  between, say, matter and vacuum as in Bonnor stars \cite{bon99, bon2}, the
  difference between these two scales becomes as large as one likes when the
  boundary approaches the position of a would-be horizon. For spherically
  symmetric spacetimes it is demonstrated explicitly in Sec.\,III of
  \cite{qbh}. In a more general setting, without spherical symmetry, this
  was traced in \cite{mn}, where the terms ``inner world'' and ``outer
  world'' were used.  Meanwhile, even if there is no sharp boundary between
  two regions of space-time, two scales and associated regions appear
  anyway. Moreover, it also makes sense to single out an intermediate region
  corresponding to the immediate vicinity of a would-be horizon. Instead of
  ``inner and outer worlds'', in our view, it is better (though also not
  quite precise) to speak of looking at a certain region ``through a
  microscope''. This expression will be explained below.

  Now, we will generalize the observations concerning the QBH metrics,
  previously made for different specific material sources (dust,
  electromagnetic and scalar fields, see \cite{1, 2}. Actually, they follow
  directly from the form (\ref{met}) of the metric.  It appears that there
  is a natural division of the whole space into three regions.

\subsection{\protect Region 1 (central), $0\le x\lesssim c$}

  Let us try to preserve a regular center in the limit $c\to 0$.
  To do so, we introduce the new coordinates
\begin{equation}
	X = x/c, \qquad T = ct/f_0, \qquad f_0 := f(0,0),      \label{sub1}
\end{equation}
  and, rewriting the metric (\ref{met}) with (\ref{A,r}) in terms
  of these new coordinates, consider $c \to 0$. The result is
\begin{equation}                                         \label{ds-lim1}
	ds^2 = (1 + X^2)dT^2 - \frac{f_0^2}{1+X^2} (dX^2 + X^2 d\Omega^2).
\end{equation}
  This limiting metric is geodesically complete, it still has a regular
  center at $X =0$, but it is not \asflat: instead, at large $X$ it approaches
  a flux-tube metric with $r(X) = \const$ and $g_{TT} \to \infty$, which
  can be characterized as a repulsive infinity for test particles.

  Even though the transformation (\ref{sub1}) applies to the whole space,
  the metric (\ref{ds-lim1}) can be seen as a result of infinitely
  stretching the neighborhood of the center (which explains the above
  mentioning of a ``microscope''). Noteworthy, it does not depend on the
  particular form of the function $f(x,c)$ and even on $f(x,0)$: the only
  remaining parameter is the constant $f_0$.

\subsection{Region 2 (intermediate, close to a quasihorizon)}

  Consider the values of $x$ such that the spherical radius $r(x,c)$ in
  (\ref{A,r}) corresponds to the horizon size of a BH obtained in a direct
  limit $c\to 0$. It makes sense to call some properly defined sphere from
  this region a ``quasihorizon''. There can be at least two reasonable
  definitions of a quasihorizon in terms of $c$ and the function $f(x,c)$
  from the ansatz (\ref{A,r}):

  {\bf First}, a quasihorizon $x=x_1$ can be defined as such a value of $x$
  that the radius $r(x_1,c)$ is simply equal to the BH horizon radius $r_h$.
  In the limit $c\to 0$ in (\ref{A,r}), we have $r(x) = f(x,0)$, and the
  horizon is located at $x=0$, therefore $r_h = f(0,0) = f_0$, and $x_1$
  should satisfy the equation $r(x_1, c) = f_0$.

  {\bf Second}, a quasihorizon can be defined using the fact that a
  horizon is a surface where $V \equiv -g^{ab} r_{,a} r_{,b} =0$ (that is,
  the gradient of the function $r(x,t)$ considered as a scalar field in the
  2D space parametrized by $x$ and $t$ becomes null). In our static case
  we have $V = A (dr/dx)^2$, accordingly a horizon corresponds to $A=0$.
  In a QBH metric, $V$ nowhere turns to zero but has a small minimum. to
  be called a quasihorizon\footnote
  	{If one uses $r$ as the radial coordinate (the usual
	Schwarzschild-like coordinate), then $-V$ coincides with the metric
	coefficient $g^{rr}$. The notion of a quasihorizon as a minimum of
	$|g^{rr}|$ was used in \cite{qbh,ext}. }
  $x = x_2$.

  Let us find $x_1$ and $x_2$ using the representation (\ref{A,r}).
  We assume that they are small in terms of $c$ (since the very existence of
  these surfaces is connected with $c\ne 0$) but still $x_{1,2} \gg c$
  (an assumption to be verified by the results). 

  Moreover, we will use the first two terms of the Taylor expansion of
  $f(x,c)$ near $x =0$, neglecting its $c$ dependence due to $c\ll x$. 
  Thus we have\footnote
	{It is the generic case, but similar results
	are obtained if the second nonvanishing term in (\ref{fn}) has
	the form $f_n x^n$, $n \in {\rm I\!N}$. In particular, the
	limiting metric for $c \to 0$ still has the form (\ref{ds-lim2}).
	}
\begin{equation}                                               \label{fn}
	f(x, c) \approx f(x) = f_0 + f_1 x + \ldots
\end{equation}
  where $f_0,\ f_1, \ldots$ are constants. 

  Now, the equation $r(x_1, c) = f_0$ is rewritten in the form
\[
	\frac{x_1}{\sqrt{x_1^2 + c^2}} (f_0 + f_1 x_1) = 1,
\]
  from which, for small $c$, it follows
\begin{equation}                                              \label{x1}
	       x_1 = a_1 c^{2/3}, \qquad a_1 = (2f_1/f_0)^{-1/3}.
\end{equation}

  To find $x_2$, we note that according to (\ref{met}) and (\ref{A,r})
  (the prime denotes $d/dx$),
\begin{equation}
	      V = A r'{}^2 = (1- x\gamma')^2,\qquad 2\gamma \equiv \ln A
\end{equation}
  therefore the condition $V' =0$, necessary for a minimum of $V$, reads
\begin{equation}                                              \label{eq-x2}
	      (x \gamma')' =0, \ \ \ {\rm or} \ \ \ x\gamma'' = - \gamma'.
\end{equation}
  Substituting $A \equiv e^{2\gamma}$ according to (\ref{A,r}) and the 
  expansion (\ref{fn}) for $f(x)$, we arrive at the equation
  $f_1/f_0 = 2c^2/x_2^3$, whence
\begin{equation}			\label{x2}
	x_2 = a_2 c^{2/3}, \qquad  
	a_2 = \big[f_1/(2f_0)]^{-1/3}.
\end{equation}
  Moreover, it can be verified that $V'' > 0$ at $x = x_2$, which confirms that 
  it is a minimum, and that possible inclusion of higher terms in the Taylor 
  expansion (\ref{fn}) does not break these results.

  Thus $x_1$ and $x_2$ coincide by order of magnitude, the difference being 
  only in the factors $a_{1,2}$. Generically $a_{1,2} = O(1)$, hence 
  $x_{1,2} = O(c^{2/3}) \gg c$, as was assumed.

  The $c$ dependence of $x_{1,2}$ prompts a possible transformation
  that could help us to approximately preserve the geometry near $x=x_{1,2}$
  in the limit $c\to 0$: let us substitute instead of $x$ and $t$
  the coordinates
\begin{equation}					\label{sub2}
       \xi = c^{-2/3} x, \qquad \tau = f_0^{-1} c^{2/3} t,     
\end{equation}
  and consider the limit $c\to 0$ in these new coordinates. The result is
\begin{equation}					    \label{ds-lim2}
       ds^2 = \xi^2 d\tau^2 - \frac{f_0^2}{\xi^2} d\xi^2 -f_0^2 d\Omega^2.
\end{equation}
  It is a pure flux-tube metric, coinciding up to a constant factor with
  the Bertotti-Robinson metric \cite{br1,br2} in agreement with Eq.\,(30)
  of \cite{qbh}. Its peculiar feature is that, although $r$ is constant,
  the metric is strongly $\xi$-dependent. Instead of a regular center, it
  contains a second-order horizon at $\xi = 0$, and the region $\xi < 0$
  beyond it is an exact copy of the region $\xi > 0$. Recall that the
  Bertotti-Robinson metric approximately describes the throat-like
  neighborhood of the extremal Reissner-Nordstr\"om (RN) metric, but, at the
  same time, it is an exact solution to the Einstein-Maxwell equations and
  can be considered without reference to the RN metric.

  The metric (\ref{ds-lim2}) at $x > 0$ can be called a ``microscope image'' of 
  the quisihorizon domain of a QBH in the following sense. While $r = \const$ 
  in the limiting metric, it changes comparatively slowly in the QBH metric 
  near $x = x_1$ or $x=x_2$: indeed, at $x = a c^{2/3}$, where $a = O(1)$, one 
  has $r'/r = O(1)$ while $\gamma' = A'/(2A) = O(c^{-2/3})$, a large quantity.
  That is, $r(x)$ is changing in this region much slower than $A(x)$.

\subsection{Region 3, $x \gtrsim x_{1,2}\gg c$.}

  In this outer region, one can simply put $c=0$ without changing the
  qualitative features of the metric, so it looks like that of a BH.

  Thus we obtain different limiting metrics preserving the main properties
  of different regions of a QBH space-time.
  For instance, there is a limiting metric preserving a regular center
  but having no spatial infinity, and some authors even describe such a
  situation saying that near the center ``there exists another world on its
  own right, with another metric'' \cite{mn}.

  Summarizing, we can say that although each kind of limiting procedure
  touches upon the whole QBH space-time, different limits preserve the
  geometric properties of different parts of space. Thus, the
  straightforward limit $c\to 0$ well preserves the geometry at large radii,
  including asymptotic flatness, while the other two transitions
  correspond to ``looking through a microscope'' at the central (1)
  and intermediate (2) regions. It is of interest that these both limiting
  metrics are universal, insensitive to the detailed behavior of the
  function $f(x,c)$. At small but nonzero $c$ these three regions overlap
  and cover the whole space. The limiting metrics can also be considered on
  their own rights as solutions to the field equations.

\subsection{QBHs with boundaries}

  So far we assumed that there is a unique metric valid in the whole space.
  This assumption can correspond either to purely field systems (to be
  exemplified in Section 3) or to matter distributions smoothly vanishing at
  large distances from the center.

  Meanwhile, many of the existing QBH models represent
  starlike objects, such as, e.g., Bonnor stars \cite{bon99} that are
  compact objects made of extremely charged dust surrounded by vacuum.     
  We can also consider such objects in a general form. Then
  the metrics are different in the inner and outer regions and are mutually
  independent up to the necessity to match them at the junction surface.

  For the whole such configuration to be close to a BH one, it is
  sufficient to require the behavior (\ref{A,r}) of the metric functions in
  the outer region only (in fact, the outer metric can be a pure BH one,
  as happens, e.g., in models using the extreme RN metric for this purpose
  \cite{bon99}). However, in a QBH the junction surface
  should be close to a would-be horizon, and this value of $x$ may be chosen
  as the small parameter of the system: the inner region is then
  $0\leq x\leq c$, where $x=0$ is the center. It then follows that at the
  junction the external function $g_{00} = A_{\rm ext} (x,c)$ should take a
  value $A_{\rm ext}(c,c) = O(c^2)$.

  The function $g_{00} = A(x,c)$ in the central region should satisfy the
  following conditions: (i) a regular center, hence $A(x,c) \approx  A_0(c)
  + A_2(c)x^2$ as $x\to 0$, and (ii) a
  smooth junction, hence $A(c,c) = A_{\rm ext}(c,c)$ and
  $A'(c,c) = A'_{\rm ext}(c,c)$ (as before, the prime means $d/dx$).
  In other respects, $A(x,c)$ is arbitrary.

  Consider the outer metric function $A(x,c)$ ($x\geq c$) in the RN form, so
  that
\begin{equation}
	A_{\rm ext} = \frac{x^{2}}{(x+m)^{2}}
\end{equation}
  Inside, at $0\leq x\leq c$, we can choose $A(x,c)$ in the form
\begin{equation}
	A = \frac{c^{2}}{[c+m+m\eta (y)]^2}, \qquad y = x/c.
\end{equation}
  The two functions (hence the metric itself) are smoothly matched at $x=c$
  provided that
\begin{equation}
	\eta (1) = 0, \qquad \frac{d\eta}{dy} (1) = -1,
\end{equation}
  while a regular center at $y=0$ requires $\eta = \eta_0 + o(y)$ at small $y$.
  This construction generalizes Bonnor's example \cite{bon99} in which
\begin{equation}
	\eta (y) = (1-y^n)/n, \qquad n \geq 2.
\end{equation}
  It is of course quite easy to invent a diversity of such metrics.
  One more good example is
\begin{equation}
	\eta = 1 - e^{(y^2-1)/2}.
\end{equation}
  After the substitutions $x=yc$, $t= T/c$, in the
  limit $c\to 0$, the metric in region 1 acquires the form
\begin{equation}
	ds^2=\frac{dT^2}{m^2(1+\eta)^2}-m^2(1+\eta)^2(dy^2+y^2d\Omega^2),	
\end{equation}
  so that the general form (\ref{met}) is preserved. Thus the limiting
  space-times of QBHs with boundaries are not unique and even contain an
  arbitrary function $\eta(y)$.

\section{QBHs with scalar and electromagnetic fields}

  So far we have been only discussing the metric properties of QBHs
  connected with the ansatz (\ref{met}). In what follows we will consider
  an example of a purely field system that can realize a full description of
  such static, \sph configurations. The Lagrangian is taken in the form
\begin{equation}                                              \label{L}
	L=\frac{1}{16\pi }[R + 2\eps (\d\chi)^{2} - F^{2}P(\chi )],
\end{equation}
  where $\eps =1$ for a normal scalar field, $\eps =-1$ for a
  phantom one, $\eps =\pm 1$, $F^2 \equiv F^{\alpha \beta }F_{\alpha\beta}$
  with $F_{\mu\nu}=\d_{\mu}A_{\nu} - \d_{\nu}A_{\mu}$ the electromagnetic
  field tensor, $A_{\mu}$ the vector potential, $\chi$ a dilatonic scalar
  field. For generality, and to provide correspondence with \cite{clem1,
  clem2}, we do not fix the sign of $P(\chi)$. The metric is assumed in the
  form (\ref{met}) and contains only one unknown function\footnote
	{More general static systems without spatial symmetry and an
	arbitrary dependence $\gamma (x^i)$ ($i = 1,2,3$) in the presence of
	scalarly and electrically charged dust have been considered in
	\cite{1, 2}, with an electric potential $\phi (x^i)$ and a dilatonic
	field $\chi (x^i)$.}
  $A(x) \equiv e^{2\gamma (x)}$. The electromagnetic field is chosen in the
  most general form compatible with spherical symmetry (only radial electric
  and magnetic fields, characterized by the charges, $q_e$ and $q_m$ are
  possible), the Maxwell-like equation leads to the squared electric field
  strength $F^{01} F_{10} = q_e^2/[r^4 P^2(\chi)]$, and the remaining field
  equations may be written in the form
\begin{eqnarray}
	x^4\gamma'' + 2x^3\gamma' &=& Ne^{2\gamma},  \label{gxx}
\\    							\label{n2}
       2\eps (x^4\chi'' + 2x^3 \chi')
     			&=& e^{2\gamma}\frac{dN}{d\chi},
\\      						\label{n}
	x^4(\gamma'{}^2 +\eps \chi'{}^2) &=& N(\chi) e^{2\gamma},
\end{eqnarray}
  where $N \equiv q_e^2/P(\chi)+ q_m^2 P(\chi)$. 

\subsection{Field configurations with a regular center}

  We are interested in \asflat\ solitonic, or particlelike configurations
  with a regular center. If, in addition, $g_{tt}\ll 1$ in some region of
  space (the gravitational redshift of emitted signals reaches large
  values at infinity where $g_{tt}=1$), we have a QBH. 

  It is in general hard to solve Eqs.\,(\ref{gxx})--(\ref{n}) with
  nontrivial $P(\chi )$ even if one of the charges, electric or magnetic, is
  zero. However, one can obtain the general result that if there is a
  regular center, it corresponds to $x=0$ and $N\sim x^{4}$ near it.

  Indeed, it can be shown that a regular center in this system can exist if
  the metric has the form (\ref{met}), and this center evidently corresponds
  to $x=0$ where $e^{2\gamma}\equiv A(x)=A_{0}+\frac{1}2 A_2 x^2 +o(x^2)$,
  where $A_0 > 0$. Substituting it into (\ref{gxx}) and (\ref{n}), we
  obtain at small $x$
\begin{equation}
	N(\chi )=\frac{3x^4 A_2 }{2A_0^2} + o(x^4),\qquad
	\eps \chi'{}^2 =\frac{3A_2 }{2A_0}+o(1).
\end{equation}
  Thus we have proved that $N=0$ at the center and, in addition, that in the
  generic case $A_2 \neq 0$ we have $\eps = {\rm sign}\, A_2$. So, with a
  normal (non-phantom) scalar field ($\eps =+1$) there is a minimum of
  $A(x)$ at the center.

  Moreover, the expression for $N(\chi)$ in terms of $P(\chi)$ shows that it
  is impossible to obtain $N=0$ if both $q_{e}\neq 0$ and $q_{m}\neq 0$. We
  thus have the following imortant result:
\medskip
\textit{Configurations with a regular center are possible with an electric
    charge (then $P(\chi) = \infty$ at the center) or with a magnetic charge
    (then $P(\chi) = 0$ at the center) and are impossible if both charges are
    nonzero.}

\medskip
  Thus, using $N(\chi)$ in the description of QBHs, we have only two
  options: \textit{either\/} $N(\chi) = q_e^2/P(\chi)$ \textit{or\/}
  $N(\chi) = q_m^2 P(\chi)$.

  It is of interest to compare the Maxwell-dilaton fields and nonlinear
  electrodynamics (NED) with the Lagrangian $L(F),\ F:=F_{\mu \nu}F^{\mu\nu}$
  as sources of gravity in what concerns the existence of self-gravitating
  configurations with a regular center \cite{kb-NED,79-ann}. In both cases
  solutions with both electric and magnetic charges and a regular center are 
  impossible (unless the NED Lagrangians are different in different parts of 
  space), but with NED the same is true for a purely electric source; on the 
  other hand, in both theories, a regular center requires a singular behavior 
  of the interaction Lagrangian: thus, in purely magnetic solutions with 
  NED a regular center requires $L(F)\to \infty $, whereas in EMD one 
   needs $P\to 0$.

\subsection{Examples}

{\bf Example 1.} Purely field configurations do not contain boundaries
  (unless there are phase transitions), so we return to the general
  metric (\ref{met}) under the condition (\ref{A,r}). Let us choose
\begin{equation}
	e^{2\gamma} \equiv A(x)=\frac{x^2 +c^2 }{(m+\sqrt{x^2 +c^2})^2 },
\end{equation}
  where $m > 0$ and $c > 0$ are constants. Such a trial function was used
  in \cite{ext, qbh}, but the material source was different there (extremal
  dust). Then at large $x$ we have $A\approx 1-2m/x+3m^2 /x^2 +...$, hence
  $m$ has the meaning of a Schwarzschild mass. Near the center
\begin{equation}
	A(x)=\frac{c^2}{(m+c)^2} + \frac{mx^2}{(m+c)^3}+\ldots ,
\end{equation}
  which confirms that the center is regular and also that with $m > 0$ the
  solution corresponds to a normal scalar field. Meanwhile, the direct
  limit $c\to 0$ leads to $A(x) = x^2/(x+m)^2$, that is, the extreme RN
  metric with a double horizon at $x=0$, $P\equiv 1$ and $\chi' \equiv 0$.

  From (\ref{gxx})--(\ref{n}) we obtain the following expressions for
  $N(\chi)$ and $\chi'$:
\begin{eqnarray}
	N(\chi) &=& \frac{mx^{4}}{z^{6}}[mx^2 +3c^2 (m+z)],
\\
	\chi'{}^2 &=& \frac{3c^2 m}{z^4(m+z)},
\end{eqnarray}
  where $z= \sqrt{x^2 +c^2}$. The function $\chi (x)$ is then obtained in
  terms of elliptic functions, and the dependence $N(\chi )$ cannot be found
  explicitly.

  As follows from the above-said, the solution contains either an electric
  or a magnetic charge and cannot contain both. In a purely electric
  solution, we have $N(\chi)=q_e^2 /P(\chi)$, hence $P\to \infty $ at the
  regular center. On the contrary, in a purely magnetic solution, we have
  $N(\chi)=q_m^2 P(\chi)$, and $P(0)=0$.

  The limiting transitions $c\to 0$ for this example of a QBH gives in
  region 1 the metric (\ref{ds-lim1}) with $f_0 = m$.
  As to region 2, at $x \gg c$ we have $f(x,c) \approx m + x$, so that in 
  the scheme of Sec. IIC we have $f_0 = m,\ f_1 =1$. Accordingly, for 
  the quasihorizon (by the two definitions in Sec. IIC) we obtain 
  $x_1 = (m/2)^{1/3}c^{2/3}$ and $x_2 = (2m)^{1/3} c^{2/3}$, and the   
  substitution (\ref{sub2}) again leads to the limiting metric (\ref{ds-lim2}).

%----------
%  and we return to Eqs.\,(22), (23) of \cite{qbh}. Equation (\ref{br})
%  corresponds to Eq.\,(30) of \cite{qbh}.
%----------

{\bf Example 2.}
  Consider the exact solitonic (particlelike) solution to the present field
  equations found in \cite{BMSh78}, see also \cite{79-ann}. We reproduce it
  here (in slightly changed notations) and discuss it from the viewpoint of
  QBHs, assuming for definiteness that there is only an electric charge and
  consequently $N(\chi) = q^2 /P(\chi)$.

  The solution is characterized by the following relations:
\begin{eqnarray}
	A(x) \equiv e^{2\gamma(x)} &=&
			\frac{(1-\delta^2)^2}{(1-\delta^2 e^{-b/x})^2},
					\label{gd}
\\
	\chi (x) &=& \chi_{0}-2\arcsin [\delta e^{-b/(2x)}], 	
\nonumber \\                   				    \label{px}
	P(\chi)
	&=& \frac{\sin^2 (\chi_{0}/2)}{\sin^2 [(\chi -\chi_{0})]},
\end{eqnarray}
  where $\sin (\chi_0/2)=\delta = m/|q|$, $b = [q^2 (1-\delta^2)]/m$.

  This solution is asymptotically flat as $x\to \infty$ and has a
  regular center at $x=0$. All its parameters are expressed in terms of the
  mass $m$ and the electric charge $q$, and $m <|q|$, so that $\delta
  <1$, while the scalar $\chi$ changes from zero at infinity to $\chi_0$
  at the center. The metric function $A(x) = e^{2\gamma}$ changes from
  $(1 - \delta^2)^2$ at the center to 1 at infinity.

  Now, let us introduce the parameter $c$ according to $1-\delta^2 = c/m$.
  The solution describes a QBH type configuration if $c/m \ll 1$.
  However, the behavior of $A(x)$ does not conform to the generic
  relations (\ref{A_c}) and (\ref{A,r}) since the function $e^{b/x}$ tends
  to zero as $x\to 0$ not in a power manner. Nevertheless, this system
  possesses the main properties of QBHs, that the size of a high-redshift
  region is not small, being of the same order of magnitude as the
  Schwarzschild radius $r_{s}=2m$.  Indeed, if we let $c\ll m$, the sphere
  $x=c$ belongs to the high-redshift region since
  $e^{\gamma}|_{x=c}=(c/m)e/(e-1)$; its radius is
  $ r|_{x=c}=m(e-1)/e\approx 0.632m$; the distance of this sphere from the
  center is $\ell (0,c)\approx 0.851m$.

  In the direct limit $c\to 0$, assuming $x \gg c$ (which corresponds to
  region 3), the metric tends to that of the extreme RN BH ($e^{\gamma} \to
  x/(x+m)$. Quite naturally, the field equations imply that in this limit
  the dilaton field becomes constant: $\chi =\chi_0 -\pi $.

  To describe region 1, we introduce the new coordinates $z = mx/c$ and
  $\tau = ct/m$, and then in the limit $c\to 0$ the metric and the dilaton
  field become
\begin{eqnarray}
	ds^2 \!&=&\! \frac{d\tau^2 }{(1{-}e^{-m/z})^2 }
		- (1{-}e^{-m/z})^2 (dz^2 + z^2 d\Omega^2),  \label{ds2-1}
\\
	\chi \!&=&\! \chi_0 - 2\arcsin e^{-m/(2z)}.
\end{eqnarray}
  This metric behaves like (\ref{ds-lim1}): it has a regular center at $z=0$
  and approaches a flux-tube form at large $z$.

  As to a possible quasihorizon, it turns out that by both definitions 1 and
  2 (see Section IIC) it now corresponds to $x \sim c$, hence it belongs to
  region 1. Therefore in this case there are only two distinct regions 1 and
  3.

\section{Conclusion}

  We have traced the general features of QBH space-times independent of 
  their specific material sources, thus generalizing a number of previous 
  observations. It turns out that one can distinguish there three typical 
  regions: the immediate vicinity of a regular centre, the quasihorizon 
  region, and the far asymptotically flat region. Different forms of limiting 
  transition $c\to 0$ convert these regions to different space-times, each of
  them being, in general, geodesically complete. Mathematically, this can be 
  viewed as examples of ``limits of space-times'' where the result of a
  limiting transition depends on how the parametersare entangled with 
  coordinates \cite{ger}.

  An important particular class of QBHs is formed by purely field 
  configurations. We have shown that, in addition to the MP or DMP systems, 
  there is an alternative way of QBH construction. To the best of our knowledge,
  it is for the first time that QBHs are obtained due to the dilaton and 
  electromagnetic fields only, without matter. These are purely field 
  configurations which do not collapse even on the very threshold of forming 
  a horizon but do not form it. This a somewhat unexpected result since it 
  cannot be given a simple interpretation in the spirit of MP systems 
  (a balance between gravitational attraction and electromagnetic repulsion). 
  It has turned out, in particular, that one of the previously found exact 
  solutions for coupled gravitational, scalar and electromagnetic fields 
  with a regular center contains a QBH at proper values of the parameters.

  It would be of interest to generalize the above results to rotating
  space-times.

\subsection*{Acknowledgments}

The work by O.Z. was supported by the subsidy allocated to Kazan Federal
University for the state assignment in the sphere of scientific activities.


\begin{thebibliography}{99}
\bibitem{qbh}
	Jos\'e P. S. Lemos and O. B. Zaslavskii, Phys. Rev. D 76, 084030
	(2007).

\bibitem{m}
	S.D. Majumdar, Phys. Rev. 72, 390 (1947).

\bibitem{p}
	A. Papapetrou, Proc. R. Ir. Acad., Sect. A 51, 191 (1947).

\bibitem{ext}
	Jos\'e P. S. Lemos and E. Weinberg, Phys. Rev. D 69, 104004 (2004).

\bibitem{1}
	K.A. Bronnikov, J.C. Fabris, R. Silveira, and O.B.
	Zaslavskii, Phys. Rev. D 89, 107501 (2014).

\bibitem{2}
	K.A. Bronnikov, J.C. Fabris, R. Silveira, and O.B.
	Zaslavskii, Gen. Rel. Grav. 46, 1775 (2014).

\bibitem{lw1}
	A. Lue and E.J. Weinberg, Phys. Rev. D 60, 084025 (1999).

\bibitem{lw2}
	A. Lue and E. Weinberg, Phys. Rev. D 61, 124003 (2000).

\bibitem{jmp}
	Jos\'e P.S. Lemos and V.T. Zanchin, J. Math. Phys. 47, 042504 (2006).

\bibitem{mn}
	R. Meinel and M. H\"{u}tten, Class. Quant. Grav. 28, 225010 (2011).

\bibitem{bon99}
	W.B. Bonnor, Class. Quantum Grav. 16, 4125 (1999).

\bibitem{bon2}
	W.B. Bonnor, Gen. Rel. Grav. 42, 1825 (2010).

\bibitem{clem1}
	G. Cl\'{e}ment, J.C. Fabris, and M.E. Rodrigues, Phys. Rev.
	D 79, 064021 (2009).{\small \ }

\bibitem{clem2}
	M. Azreg-Ainou, G. Cl\'{e}ment, J.C. Fabris, and M.E. Rodrigues,
	Phys. Rev 83, 124001 (2011); arXiv: 1102.4093.

\bibitem{br1}
	B. Bertotti, Phys. Rev. 116, 1331 (1959).

\bibitem{br2}
	I. Robinson, Bull. Acad. Pol. Sci. 7, 351 (1959).

\bibitem{kb-NED}
	K.A. Bronnikov, Phys. Rev. D 63, 044005 (2001); gr-qc/0006014.

\bibitem{79-ann}
	K.A. Bronnikov, V.N. Melnikov, G.N. Shikin, and K.P. Staniukovich,
	Ann. Phys. (NY) 118 (1), 84 (1979).

\bibitem{BMSh78}
	K.A. Bronnikov, V.N. Melnikov, and G.N. Shikin, Izvestiya Vuzov
	SSSR, Fiz., No. 11, 69 (1978); Russ. Phys. J. 21 (11), 1443 (1978).

\bibitem{ger}
	R. Geroch, Commun. Math. Phys. 13, 180 (1969).

\end{thebibliography}
\end{document}